\documentclass[preprintnumbers,amsmath,amssymb,floatfix,12pt,superscriptaddress,nofootinbib]{revtex4}
\usepackage{graphicx}
\usepackage{epsfig}
\usepackage{bm}
\usepackage{amsfonts}

\begin{document}

\title{Gravitational collapse of radiation fluid in higher dimensions}

\author{Mubasher Jamil}
\email{mjamil@camp.nust.edu.pk} \affiliation{Center for Advanced
Mathematics and Physics, National University of Sciences and
Technology, Rawalpindi, 46000, Pakistan}

\author{M. Umar Farooq}
\email{m_ufarooq@yahoo.com} \affiliation{Center for Advanced
Mathematics and Physics, National University of Sciences and
Technology, Rawalpindi, 46000, Pakistan}

\begin{abstract}
\textbf{Abstract:} We examine the problem of the gravitational
collapse using higher dimensional Husain spacetime for the null
fluid. The equations of state chosen to solve the field equations
contain linear, quadratic and arbitrary powers of the radial
parameter. The resulting mass evolution is discussed for each case.
\end{abstract}
\maketitle


\newpage
\section{Introduction}
The problem of gravitational collapse of massive bodies is
fascinating and a long standing one in general relativity. This
problem arises since general relativity predicts vanishing pressure
gradients against daunting gravitational forces in massive objects
(of the order of solar masses). Generally, spherically symmetrical
object of mass $M$ and radius $r$ related by $r\sim2M$ (in units
$c=G=1$), undergoes unrestricted collapse \cite{penrose}. This
process may not be spherically symmetrical if the collapsing object
possesses angular momentum. The theory of gravitational collapse has
been widely studied. Examples are: Collapse of a massive dust cloud
\cite{snyder}, charged perfect fluid sphere \cite{mashoon}, rotating
massive body \cite{wagoner,cohen}, role of bulk viscosity during
collapse \cite{nakia,herrera}, collapse in the background of
cosmological constant \cite{shapiro}, thin spherical shell of dust
\cite{israel}, collapse of homogeneous scalar fields \cite{roberto}
and collapse in higher dimensional spacetimes \cite{bizon,goswami}.

Almost all the gravitational collapse models lead to the formation
of spacetime singularities generally hidden by one or more horizons.
A singularity is a region where invariants like Kretschman scalar
and curvature scalar diverge. Numerical simulations of gravitational
collapse of spheroids show that if the collapsing spheroid is
sufficiently compact, the singularities are hidden inside the event
horizon while they become naked (devoid of event horizon) if the
spheroid is sufficiently large \cite{shapiro1}. However, there are
some models in which the formation of singularity is avoided e.g. if
the collapsing star radiates all the matter \cite{torres}. Another
such model is that of a `regular phantom black hole' which contains
Schwarzschild like causal structure and the singularity is replaced
by the de Sitter infinity \cite{bronn1}.

There has been huge interest in naked singularities, although their
existence is not very clear and these are prohibited by the cosmic
censorship hypothesis \cite{patel}. The existence and formation of
naked singularities has been suggested for the gravitational
collapse in self-similar spacetimes \cite{lake}. The visibility of a
singularity is possible if there exists a null geodesic emanating
from the singularity. Then it requires the existence of families of
future directed non-spacelike curves which emanate from the vicinity
of the singularity \cite{joshi}. The observation of these
non-spacelike curves will give sufficient information about the
singularity itself. Such a mysterious singularity can also be
observable if sufficiently strong shearing effects near the
singularity delay the formation of the event horizon \cite{joshi1}.
Another such possibility is that of a black hole accreting phantom
energy which results in its `evaporation' and leading to a naked
singularity \cite{babichev,jamil,jamil1,jamil2,jamil3}.
Astrophysically, the phenomenon of gravitational collapse is
manifested in the form of a gamma ray burst in which a super-giant
star explodes and releases immense heat flux while the stellar core
collapses to form a black hole remanent \cite{zhe}.

The problem of spherical collapse of a null fluid has been studied
earlier by Husain \cite{viqar} and recently extended to higher
dimensions by Debnath et al \cite{debnath}. We here investigate the
same problem using equations of state which are more general than
the barotropic EoS $p=\omega\rho$. These EoS yield interesting
behaviors for the evolution of mass of the black hole.

\section{Modeling of system}

We assume an ($n+2$)-dimensional spherically symmetric Husain
spacetime given by \cite{debnath}
\begin{equation}
ds^{2}=-\left(1-\frac{m(v,r)}{r^{n-1}}\right)dv^{2}+2dvdr+r^{2}d\Omega
_{n}^{2},
\end{equation}
where the radial coordinate is restricted in the range $0< r<\infty$
and the advanced null coordinate $v=t-r$ with $-\infty\leq
v\leq\infty$ is called the Eddington coordinate. Here we have the
energy momentum tensor with two components: null radiation fluid and
the matter fluid i.e.
\begin{equation}
T_{\mu \nu }=T_{\mu \nu }^{(n)}+T_{\mu \nu }^{(m)},
\end{equation}
where
\begin{equation}
T_{\mu \nu }^{(n)} =\sigma l_{\mu }l_{\nu },
\end{equation}
and
\begin{equation}
 T_{\mu \nu }^{(m)}
=(\rho +p)(l_{\mu }\eta _{\nu }+l_{\nu }\eta _{\mu })+pg_{\mu\nu}.
\end{equation}
Everywhere in this paper, all Greek indices range from 1 to $n+2$.
Here $l^\mu=(1,0,0,...,0)$, and
$\eta_\mu=\left(\frac{1}{2}\left(1-\frac{m}{r^{n-1}}\right),-1,0,...,0\right)$,
are future-like Null vectors, satisfying $l_\lambda
l^\lambda=\eta_\lambda\eta^\lambda=0$, and
$l_\lambda\eta^\lambda=-1$. Also $\sigma$ is the is the energy
density corresponding to the Vaidya null direction. We require the
energy momentum tensor to satisfy the energy conditions given by (a)
Weak and strong energy conditions are: $\sigma >0,\ \rho \geq 0,\
p\geq 0$. (b) Dominant energy condition (DEC) is: $\sigma >0,\ \rho
\geq p,\ p\geq 0$.

The Einstein field equations are
\begin{equation}
G_{\mu \nu }= T_{\mu \nu },
\end{equation}
for the metric (1) with matter field having stress-energy tensor
given by
\begin{eqnarray}
\rho &=&\frac{nm^{\prime }}{2r^{n}},\\ p&=&-\frac{m^{\prime \prime
}}{2r^{n-1}}, \\  \sigma &=&\frac{n\dot{m}}{2r^{n}}.
\end{eqnarray}
Here prime $'$ and overdot $.$ denote differentiation with respect
to the parameters $r$ and $v$ respectively. For the positive
definiteness of $\rho$, $p$ and $\sigma$, we require
\begin{equation}
(a)\ m^{\prime }\geq 0,\ m^{\prime \prime }\leq 0 \ and \ \ (b) \
\dot{m}>0.
\end{equation}
We take the following cases of equations of state to solve the field
equations (6) - (8):
\begin{enumerate}
  \item $p(v)=-Ar+B\rho(v)$,
  \item $p(v)=(Dr+Er^2)\rho(v)$,
  \item $p(v)=Cr^k\rho(v)$.
\end{enumerate}
Here $A$, $B$, $C$, $D$ and $E$ are arbitrary constants independent
of both $r$ and $v$. Note that the barotropic EoS $p=\omega\rho$,
represents an asymptotically flat spacetime if the parameter
$\omega$ is constrained by $\frac{1}{2}<\omega\leq 1$, while
$\omega=1$ represents a charged Vaidya solution \cite{viqar}.

\subsection{Case-1: EoS Linear in $r$}

We consider an EoS which is linear in variable $r$ and is given by
\begin{equation}
p(v)=-Ar+B\rho (v),
\end{equation}
Using Eqs. (6), (7) and (10), we get
\begin{equation}
m^{\prime \prime }(r,v)=2 Ar^{n}-\frac{n B m^{\prime }(r,v)%
}{r}.
\end{equation}
Solving Eq. (11), we obtain
\begin{equation}
m(r,v)=C_{1}(v)+C_{2}(v)\frac{r^{1-Bn}}{1-Bn}+\frac{2Ar^{2+n}}{(2+n)(1+n+Bn)%
},
\end{equation}
where $C_1(v)$ and $C_2(v)$ are arbitrary functions of time $v$.
Differentiating Eq. (12), w.r.t $r$ yields
\begin{equation}
m^{\prime }(r,v)=C_{2}(v)r^{-Bn}+\frac{2Ar^{n+1}}{1+n+Bn},
\end{equation}
while second differentiation gives
\begin{equation}
m^{\prime \prime
}(r,v)=-C_{2}(v)Bnr^{-1-Bn}+\frac{2A(n+1)r^{n}}{1+n+Bn}.
\end{equation}
Also differentiation of Eq. (12) w.r.t $v$ gives
\begin{equation}
\dot{m}(r,v)=\dot{C}_{1}(v)+\dot{C}_{2}(v)\frac{r^{1-Bn}}{1-Bn}.
\end{equation}
Now using Eq. (13) $m^{\prime }\geq 0$ $\Rightarrow$ $\frac{
2Ar^{n+1}}{1+n+Bn}+r^{-Bn}C_{2}(v)\geq 0$. It yields
\begin{equation}
r\geq \left[ -\frac{C_{2}(1+n+Bn)}{2A}\right]^{\frac{1}{1+n+Bn}}.
\end{equation}
Now $-\frac{2A}{C_{2}(1+n+Bn)}\geq0$ if either $A>0$ and
$C_{2}(1+n+Bn)<0$ or vice-versa. The later quantity yields $ C_{2}>
0$ and $1+n+Bn<0$ and vice versa. Also $m^{\prime \prime }\leq 0$
implies
\begin{equation}
\frac{2A(n+1)r^{n}}{1+n+Bn}-Bnr^{-1-Bn}C_{2}(v)\leq 0,
\end{equation}
which gives
\begin{equation}
r^{1+n+Bn}\leq Bn\frac{(1+n+Bn)}{2A(1+n)}.
\end{equation}
Further, $\dot{m}>0$ implies
\begin{equation}
\dot{C}_{1}(v)+\dot{C}_{2}(v)\frac{r^{1-Bn}}{1-Bn}> 0,
\end{equation}
or
\begin{equation}
\frac{ \dot{C}_{1}}{\dot{C}_{2}}>\frac{r^{1-Bn}}{Bn-1}.
\end{equation}
Now horizon of the metric is obtained by
$1-\frac{m(r,v)}{r^{n-1}}=0$. It implies $m(r,v)=r^{n-1}$ which
further yields
\begin{equation}
C_{1}(v)+C_{2}(v)\frac{r^{1-Bn}}{
1-Bn}+\frac{2Ar^{2+n}}{(2+n)(1+n+Bn)}=r^{n-1},
\end{equation}
which is an algebraic equation in $r$.

In the graphs to follow, we plot the evolution of a black hole mass
by considering different models (i.e. cases 1 to 3) resulting from
the equations of state of a null fluid. Our graphs result from
different choices of functions $C_i$, $i=1...6$ which are chosen
quite arbitrarily including polynomial, trigonometric and the
exponential functions, and exhibit various behaviors for the mass
parameter $m$. These functions $C_i$, then lead to increasing,
decreasing or fluctuating manner of mass.

In figure 1, we have chosen $C_1(v)=e^{v^2}$ and $C_2(v)=\sin v^3$.
The constant parameters are fixed at $A=3$, $B=-8$ and $n=4$. It is
shown that mass increases in steps as $r$ increases for $v>0$
whereas for $v<0$, the mass increases as $r$ decreases. In figure 2,
the functions are chosen as $C_1(v)=e^{v^2}$ and $C_2(v)=\sin v^2$;
while the constants are taken to be $A=-3$, $B=-2$ and $n=4$. The
graph shows that this model is similar to the previous one except
for the symmetry in $v$ i.e. the mass increases with the increases
in radial coordinate $r$.

\subsection{Case-2: EoS quadratic in $r$}

We now take another EoS which is quadratic in $r$ given by
\begin{equation}
p(v)=(Dr+Er^{2})\rho (v).
\end{equation}
The governing equation is
\begin{equation}
m^{\prime \prime }(r,v)=-n\left(D+Er\right)m^{\prime }(r,v).
\end{equation}
Solving Eq. (23), we obtain
\begin{equation}
m(r,v)=C_{3}(v)+C_{4}(v)\exp \left(\frac{D^{2}n}{2E
}\right)\sqrt{\frac{2}{\pi E}}\int\limits_{0}^{\sqrt{\frac{n}{2E
}}(D+Er)}e^{-z^2}dz.
\end{equation}
Also differentiation of Eq. (24) w.r.t $r$ gives
\begin{equation}
m^{\prime }(r,v)=C_{4}(v)\exp \left(-\frac{1}{2}nr(2D+Er)\right).
\end{equation}
Substitution of (25) in (23) gives
\begin{equation}
m^{\prime \prime }(r,v)=-nC_{4}(v)(D+Er)\exp
\left(-\frac{1}{2}nr(2D+E r)\right).
\end{equation}
Differentiation of Eq. (24) w.r.t $v$ results
\begin{equation}
\dot{m}(r,v)=\dot{C}_{3}(v)+\dot{C}_{4}(v)\exp
\left(\frac{D^{2}n}{2E }\right)\sqrt{\frac{2}{\pi
E}}\int\limits_{0}^{\sqrt{\frac{n}{2E }}(D+Er)}e^{-z^2}dz.
\end{equation}
Now $m^{\prime }(r,v)\geq 0$ $\Rightarrow C_{4}(v)\exp
(-\frac{1}{2}nr(2D+Er))\geq 0$ $\Rightarrow C_{4}(v)\geq 0$.
Further, $m^{\prime \prime }(r,v)\leq 0 \Rightarrow -n(D+E
r)C_{4}(v)\leq 0$, when $n(D+Er)\geq0$. Also $\dot{m}(r,v)>0$
implies
\begin{equation}
\frac{\dot{C}_{3}(v)}{\dot{C}_{4}(v)}>-\exp \left(\frac{D^{2}n}{2E
}\right)\sqrt{\frac{2}{\pi E}}\int\limits_{0}^{\sqrt{\frac{n}{2E
}}(D+Er)}e^{-z^2}dz.
\end{equation}
Now to calculate the horizon we take $m(r,v)=r^{n-1}$ which gives
\begin{equation}
C_{3}(v)+C_{4}(v)\exp \left(\frac{D^{2}n}{2E
}\right)\sqrt{\frac{2}{\pi E}}\int\limits_{0}^{\sqrt{\frac{n}{2E
}}(D+Er)}e^{-z^2}dz=r^{n-1}.
\end{equation}
In figure 3, we have chosen $C_3(v)=e^{v^3}$ and $C_4(v)=\cot v^2$.
The constant parameters are fixed at $D=2$, $E=3$ and $n=6$. Here
the mass $m$ possesses symmetry about $v=0$. The mass eventually
decreases for large $v$. In figure 4, the functions are chosen as
$C_3(v)=e^{v}$ and $C_4(v)=\csc v^2$ with the same choice of
constants as in Fig. 3. The mass decreases when $v$ increases except
for the singularity at $v=0$.

\subsection{Case-3: EoS with arbitrary power in $r$}

Let us now take the EoS
\begin{equation}
p(v)=Cr^{k}\rho (v).
\end{equation}
Here $C$ and $k$ are arbitrary constants. The governing differential
equation is given by
\begin{equation}
m^{\prime \prime }(r,v)=-nCr^{k-1}m^{\prime }(r,v).
\end{equation}
The solution of the above equation is
\begin{equation}
m(r,v)=C_{5}(v)-\frac{r}{k}C_{6}(v)\left(\frac{Cnr^{k}}{k}\right)^{-\frac{1}{k}}\Gamma
\left( \frac{1}{k},\frac{Cnr^{k}}{k}\right),
\end{equation}
where we have an incomplete Gamma function given by
\begin{equation}
\Gamma
\left(\frac{1}{k},\frac{Cnr^{k}}{k}\right)=\int\limits_{\frac{Cnr^{k}}{k}}^\infty
x^{\frac{1}{k}-1}e^{-x}dx.
\end{equation}
Differentiation of Eq. (32) w.r.t $r$, we have
\begin{equation}
m^{\prime }(r,v)=C_{6}(v)\exp \left(-\frac{Cnr^{k}}{k}\right).
\end{equation}
Again differentiating yields
\begin{equation}
m^{\prime \prime }(r,v)=-Cnr^{k-1}C_{6}(v)\exp
\left(-\frac{Cnr^{k}}{k}\right).
\end{equation}
Further differentiation w.r.t $v$ gives
\begin{equation}
\dot{m}(r,v)=\dot{C}_{5}(v)-\frac{r}{k}\dot{C}_{6}(t)\left(\frac{Cnr^{k}}{
k}\right)^{-\frac{1}{k}}\Gamma
\left(\frac{1}{k},\frac{Cnr^{k}}{k}\right),
\end{equation}
Now $m^{\prime }(r,v)\geq 0$ implies $C_6(v)\geq
0$. Also $m^{\prime \prime }(r,v)\leq 0$ implies $
-Cnr^{k-1}C_{6}(v)\exp (-\frac{Cnr^{k}}{k})\leq 0$, hence
$Cnr^{k-1}C_{6}(v)\geq0$. Further $\dot{m}(r,v)>0$ implies
\begin{equation}
\frac{ \dot{C}_{5}(v)}{\dot{C}_{6}(v)}>
\frac{r}{k}\left(\frac{Cnr^{k}}{k}\right)^{- \frac{1}{k}}\Gamma
\left(\frac{1}{k},\frac{Cnr^{k}}{k}\right).
\end{equation}
Now horizon of the spacetime is obtained by solving
\begin{equation}
C_{5}(v)-\frac{C_{6}(v)r}{k}\left[\left(
\frac{Cnr^{k}}{k}\right)^{-\frac{1}{k}}\Gamma\left
(\frac{1}{k},\frac{Cnr^{k}}{k}\right)\right] =r^{n-1}.
\end{equation}
In figure 5, we have chosen $C_5(v)=e^{v^2}$ and $C_6(v)=v^4$. The
constant parameters are fixed at $c=2$, $k=6$ and $n=3$. In figure
6, the functions are chosen as $C_5(v)=e^{v^3}$ and $C_6(v)=v^3$
with the same choice of constants as in Fig. 5. The mass increases
as $v$ increases in both cases. Thus the accretion of the null fluid
results in the increase in mass of the black hole.

Also note that the dominant energy condition $\rho\geq p$ implies
\begin{equation}
r\geq -\frac{ nm^{\prime }}{m^{\prime \prime }}.
\end{equation}
which is a general expression. For $p=-Ar+B\rho $, it gives
\begin{equation}
r\geq -n\left[\frac{2Ar^{n+1}+c_{1}r^{-Bn}(1+n+Bn)}{
2A(n+1)r^{n}-c_{1}Bnr^{-1-Bn}(1+n+Bn)}\right].
\end{equation}
For $p=(Dr+Er^{2})\rho (t)$, the condition Eq. (39) leads to
\begin{equation}
r\geq\frac{1}{D+Er}.
\end{equation}
It further gives $(D+Er)r\geq 1\Rightarrow r\geq 1$ and $
(D+Er)r\geq 1$. The later yields $ r\geq \frac{1-D}{E}$.

In the third case, $p=Cr^{k}\rho (t)$ the DEC implies
\begin{equation}
r\geq \frac{1}{Cr^{k-1}}\Rightarrow r\geq
\left(\frac{1}{C}\right)^{\frac{1}{k}}.
\end{equation}

\section{Conclusion}

In this paper, we have investigated the gravitational collapse model
of higher dimensional Husain spacetime. We have obtained three
different expressions of mass in the corresponding three cases.
These expressions contain certain functions $C_i$ which needs to be
chosen arbitrarily since no boundary conditions are imposed on the
governing dynamical equations. However our choices of these
functions lead to some interesting results: In cases 1 and 3, the
mass of black hole is increasing due to accretion of null fluid.
These solutions physically describe the \textit{inward (ingoing)
Husain spacetime} \cite{poisson}. However the solutions obtained in
case-2 describe the \textit{outward (outgoing) Husain spacetime}
since the mass is decreasing. Our solutions also satisfy the weak
and dominant energy conditions which are necessarily satisfied in
the classical gravity. Moreover, the equations of state chosen here,
are generalizations of the previously used ones in \cite{viqar} and
hence give a much deeper understanding of the process.

\subsubsection*{Acknowledgment}
We would like to thank Viqar Husain, Ujjal Debnath and Alberto Saa
for useful correspondence related to this work.

\pagebreak
\newpage
\begin{figure}
\includegraphics{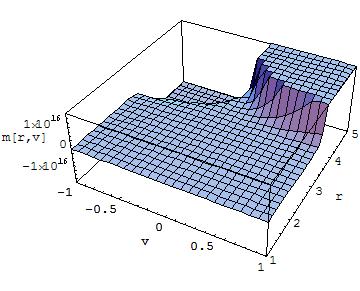}\\
\caption{The mass $m(r,v)$ is plotted against $r$ and $v$ for the
choice of functions $C_1$ and $C_2$.}
\end{figure}
\begin{figure}
\includegraphics{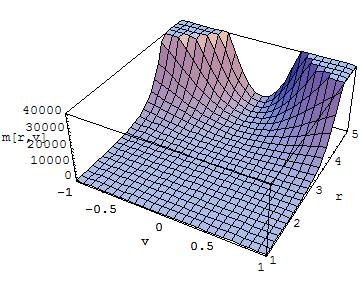}\\
\caption{The mass $m(r,v)$ is plotted against $r$ and $v$ for the
choice of functions $C_1$ and $C_2$.}
\end{figure}
\begin{figure}
\includegraphics{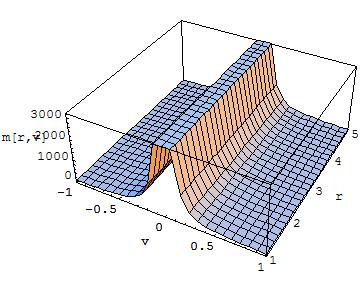}\\
\caption{The mass $m(r,v)$ is plotted against $r$ and $v$ for the
choice of functions $C_3$ and $C_4$.}
\end{figure}
\begin{figure}
\includegraphics{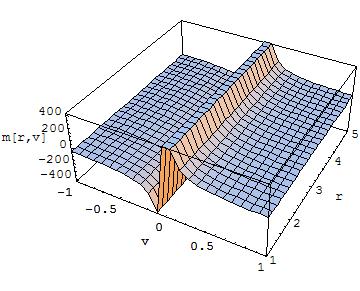}\\
\caption{The mass $m(r,v)$ is plotted against $r$ and $v$ for the
choice of functions $C_3$ and $C_4$.}
\end{figure}
\begin{figure}
\includegraphics{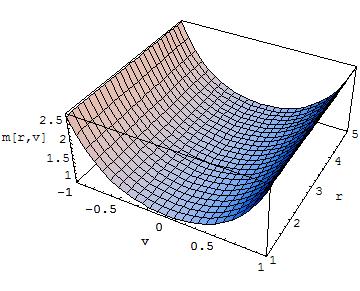}\\
\caption{The mass $m(r,v)$ is plotted against $r$ and $v$ for the
choice of functions $C_5$ and $C_6$.}
\end{figure}
\begin{figure}
\includegraphics{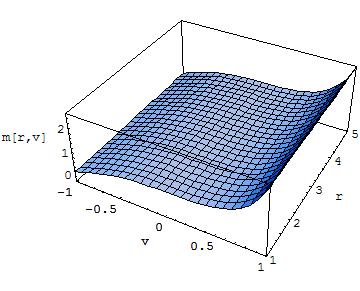}\\
\caption{The mass $m(r,v)$ is plotted against $r$ and $v$ for the
choice of functions $C_5$ and $C_6$.}
\end{figure}
\end{document}